\documentclass[final]{svjour2}
\usepackage{graphicx}
\graphicspath{    
{images/}
} 
\usepackage{rotating}
\usepackage{amssymb}
\usepackage{mathptmx}
\usepackage[numbers]{natbib}
\makeatletter
\journalname{Journal of Low Temperature Physics}

\bibpunct{}{}{,}{s}{}{,}

\begin{document}

\newcommand{\hdblarrow}{H\makebox[0.9ex][l]{$\downdownarrows$}-}
\title{Approaches to the Optimal Nonlinear Analysis of Microcalorimeter Pulses}

\author{J.W.~Fowler  \and C.G.~Pappas  \and B.K.~Alpert, \and W.B.~Doriese, \and G.C.~O'Neil, \and J.N.~Ullom, \and D.S.~Swetz}

\institute{Quantum Sensors Group, NIST Boulder Laboratories, 325 Broadway, MS 687.08, Boulder, Colorado 80305, USA. Tel: +1 303-497-3990.\\
\email{joe.fowler@nist.gov}}
 
\vspace{-1cm}
 \date{Draft of: \today}
\maketitle
\vspace{-5mm}

\begin{abstract}

We consider how to analyze microcalorimeter pulses for quantities that are nonlinear in the data, while preserving the signal-to-noise advantages of linear optimal filtering. We successfully apply our chosen approach to compute the electrothermal feedback energy deficit (the ``Joule energy'')  of a pulse, which has been proposed as a linear estimator of the deposited photon energy.


\end{abstract}

\newcommand{\Rshunt}{\ensuremath{R_\mathrm{sh}}}
\newcommand{\Ibias}{\ensuremath{I_\mathrm{bias}}}
\newcommand{\Rq}{\ensuremath{R_\mathrm{q}}}
\newcommand{\Iq}{\ensuremath{I_\mathrm{q}}}
\newcommand{\dif}{\ensuremath{\mathrm{d}}}
\newcommand{\Ej}{\ensuremath{E_\mathrm{Joule}}}

\section{Overview of the Problem}

\begin{figure}[tbp]
\begin{center}
\includegraphics[width=\linewidth, keepaspectratio]{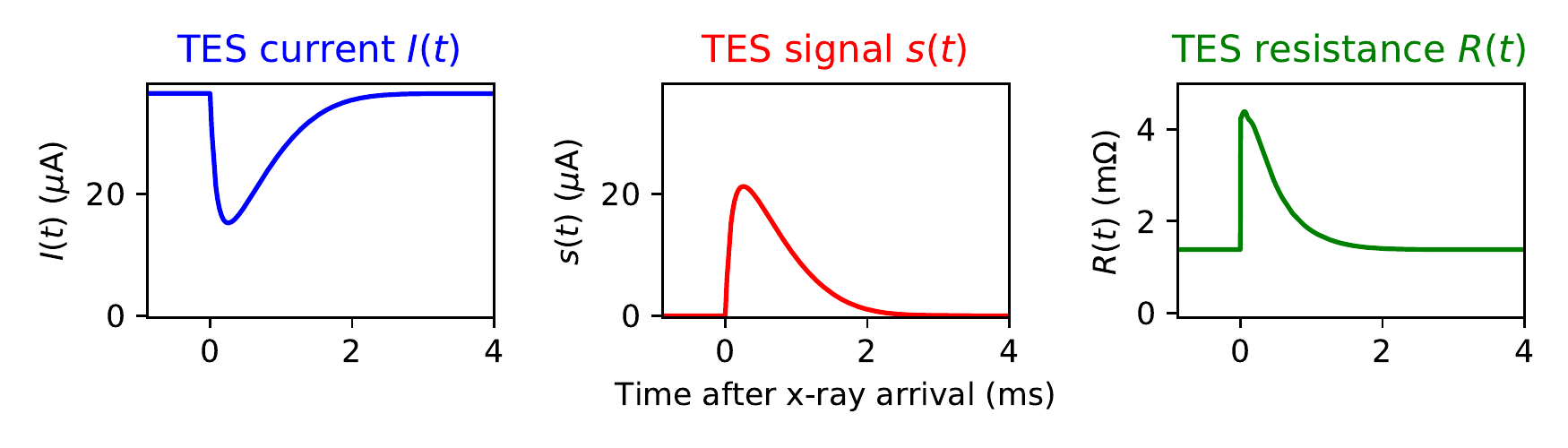}
\caption{\label{fig:signal}
A typical signal pulse from a TES detector. \emph{Left:} the detector current $I(t)$ is a transient \emph{reduction} in current from its quiescent level \Iq. \emph{Center:} the same pulse in terms of the ``signal'' $s(t)\equiv \Iq - I(t)$, as used throughout this paper.
\emph{Right:} the TES resistance, estimated from $s(t)$ and measured electrical parameters. (Color figure online.)}
\vspace{-5mm}
\end{center}
\end{figure}

A transition-edge sensor (TES) microcalorimeter responds electrically to the absorption of an x-ray photon because the photon heats the sensor, which increases its resistance (Fig.~\ref{fig:signal}). The TES is electrically biased with negative feedback to balance it in the superconducting transition, where the resistance is very sensitive to temperature changes. The same phase transition that produces this exquisite sensitivity also produces an appreciable nonlinearity, however, so the TES signal is not simply proportional to the deposited energy. In our experience, this nonlinearity is present at all energies, even far from detector saturation.

Usually, the analysis of TES data employs a technique known as \emph{optimal filtering} to achieve high energy resolution.\cite{fowler_ppp, alpert2013} Many consecutive samples of the TES bias current are recorded over a period of some milliseconds before and after the x-ray absorption to track the transient, pulsed decrease in this current. Optimal filtering prescribes a particular linear combination of these $N$ samples, which estimates the pulse size. This weighting is statistically optimal, under certain conditions: that noise is Gaussian, additive, and stationary at any signal level; that the pulses are always the same shape, regardless of the absorbed energy; that pulses are transient departures from a strictly constant quiescent, or ``baseline,'' current level; and that no photon arrives soon enough after another for either to have an appreciable effect on the other's pulse signal. Although none of these conditions are strictly obeyed in real data, the method nevertheless works very well, and certain corrections can reduce the impact of small systematic errors that follow from the violation of the assumptions.\cite{fowler_ppp}

Optimal filtering estimates pulse \emph{sizes} well but does not address the problem that pulse size is not proportional to pulse energy.  Extensive effort has been devoted in recent years to generalize optimal filtering when a range of pulse shapes must be accommodated\cite{smith2009,fixsen2014,shank2014,busch2016,yan2016,fowlertangent2017} or to apply it to a time-series such as TES resistance for improved energy linearity.\cite{bandler2006,sjlee2015,peille2016}
In a companion paper,\cite{pappas17} we explore a nonlinear transformation of the data which we call the \emph{Joule energy} of a pulse, also known as the electrothermal feedback energy.\cite{irwin_hilton} It measures the deficit in the Joule heating in the TES that occurs during (and because of) a pulse. Though this Joule energy accounts for only a portion of the total x-ray energy deposited in the sensor (typically well more than half), it is nevertheless very nearly  \emph{proportional} to the total energy over a broad range of energies so long as the cryogenic bath temperature and TES bias voltage are held constant. Therefore, we ask whether we might estimate the Joule energy on a pulse-by-pulse basis as a potential improvement upon the optimal filter, which estimates only pulse sizes.

Estimation of Joule energy by the straightforward computation of the relevant time integrals from the noisy signal data, however, produces a very noisy result. In this paper, we explore how to estimate the time integrals in a statistically optimal fashion, analogous to optimal filtering but with the additional benefit of much improved energy linearity. The method described here is a first attempt to test whether optimal Joule-energy estimation is possible.

\section{The Joule Energy of a Pulse}

By the \emph{Joule energy} of a pulse, we mean the time integral of the reduction in the Joule power ($I^2R$) dissipated in the TES, compared to that dissipated in the TES's quiescent state. 
We model the bias circuit as a constant current \Ibias\ split between a shunt resistor of resistance \Rshunt\ and an inductance $L$ in series with $R(t)$, the TES resistance. Let \Rq\ be the quiescent value of $R(t)$. The TES quiescent current is $\Iq = \Ibias \Rshunt/(\Rq+\Rshunt)$. The Joule power delivered to the TES is
\[
P(t) = I(t) V(t) = 
\Rshunt[\Ibias I(t) - I^2(t)] - L I(t) \dif{I}/\dif{t},
\]
where $I(t)$ and $V(t)$ are the TES current and voltage.
The time integral of the  quiescent power minus the dynamic power $P(t)$ defines the Joule energy:
\[
\Ej \equiv \int_{t_0}^{t_f}\,\dif t \left\{ \Iq^2 R_\mathrm{q} - \Rshunt[\Ibias I(t) - I^2(t)] + L I(t) \dif{I}/\dif{t} \right\}.
\]
Here, $t_0$ and $t_f$ represent times before the photon arrival and well after the pulse ends, respectively.
The last term integrates to zero, because  $I(t)=\Iq$ both at the beginning and end of the interval (i.e., there is zero net change in power stored in the inductor). We re-express $I(t)$ in terms of the positive-going ``signal'' (Fig.~\ref{fig:signal}), the reduction in TES current from its quiescent value, $s(t)\equiv \Iq - I(t)$, to find:
\begin{equation} \label{eq:EJoule}
\Ej = \Rshunt(\Ibias - 2\Iq)\int_{t_0}^{t_f}\,\dif t\ s(t) + \Rshunt\int_{t_0}^{t_f}\,\dif t\ s^2(t).
\end{equation}
Thus, the Joule energy is a linear combination of the time integrals of $s(t)$ and of its square through the duration of the pulse. The weights of the integrals are given by certain constant electrical parameters of the circuit.


\begin{figure}[htbp]
\begin{center}
\includegraphics[width=\linewidth, keepaspectratio]{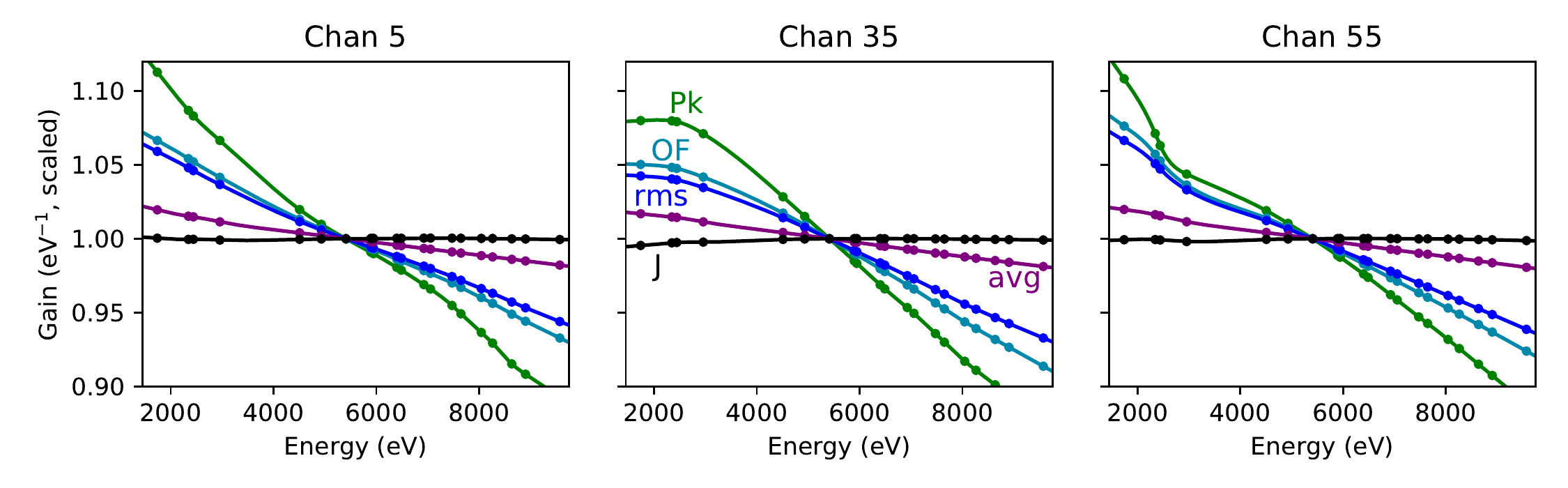}
\caption{\label{fig:gains}
The ``gain'' $G$ (estimator per unit energy) versus energy for five pulse-size estimators. The three panels show different TESs each with its own distinct $G$ vs $E$ pattern. Points show the mean $G$ for twenty calibration lines, with splines connecting the points. From least to most linear, the estimators are: the pulse peak value (\emph{Pk}); the result of optimal filtering (\emph{OF}); the root-mean-square signal-above-baseline (\emph{rms}); the mean signal-above-baseline (\emph{avg}); and the Joule energy (\emph{J}). All estimators are normalized to $G=1$ at Cr K$\alpha$ (5415\,eV).
(Color figure online.)}
\end{center}
\end{figure}

We have argued\cite{pappas17} that under typical operating conditions, for a fixed bath temperature and bias current, the Joule energy of a pulse should be nearly proportional to the energy of the photon that caused it. We find this to be true empirically over a broad range of measurements. Fig.~\ref{fig:gains} shows five estimators of energy for each of three sensors. The Joule energy is most nearly linear by far: the gain for optimal filtering falls by some 15\,\% across the 2--9\,keV range; the Joule energy's gain varies by no more than 0.5\,\% over the same range. The pulse average is the most linear of the standard estimators but is far noisier than the optimal filter.


\section{Optimal Nonlinear Estimates of Joule Energy}

Although Joule energy is proportional to the photon energy, it is unfortunately \emph{not} linear in the TES current measurements, owing to the $s^2(t)$ term in Equation~\ref{eq:EJoule}. We define an energy estimator $J$, proportional to the Joule energy:
\begin{equation}\label{eq:J}
J \equiv \xi\Ej =  \lambda\int\,\dif t\ s(t) + \sigma\int\,\dif t\ s^2(t),
\end{equation}
where $\xi$ is the unknown ratio of photon to Joule energy, and $\lambda$ and $\sigma$ are the unknown weights of the linear and squared signal terms. We can read the ratios $\lambda/\xi$ and $\lambda/\sigma$ by comparison of Eq.~\ref{eq:J} to Eq.~\ref{eq:EJoule}, but they depend on circuit parameters that may be difficult to measure to the desired level of precision. Worse, the overall scale factor $\xi$ can be computed only from a complete electro-thermal model of the TES. 
We choose $\lambda$ and $\sigma$ of the linear and quadratic signal integrals to yield the best linearity in the data at hand via a simple least-squares fit. Additionally, the $\sigma$ so chosen can correct for leading-order nonlinearity in $\Ej$.
(Other procedures are first used to estimate pulse energies, but weight selection should be straightforward even in the absence of prior energy estimates.) Once $\lambda$ and $\sigma$ are chosen, we can turn to the nonlinear problem of estimating $J$ from $s(t)$.


The signal $s(t)$ is noisy; if we perform the integrals of Eq.~\ref{eq:J}  numerically on $s(t)$, they yield a noisy estimate of $J$. A low-noise estimator can be designed, however, if we can identify---within the $N$-dimensional space that represents all possible records---a one-dimensional curve where all noise-free 
pulses would lie. By confining all pulses to such a curve, we can eliminate much of the noise intrinsic to the measurement, much as optimal filtering does. We construct this curve from a large set of ``training data'' in two steps, first identifying a low-dimensional \emph{linear} subspace that (approximately) contains the curve.

\begin{figure}[htbp]
\begin{center}
\includegraphics[width=\linewidth, keepaspectratio]{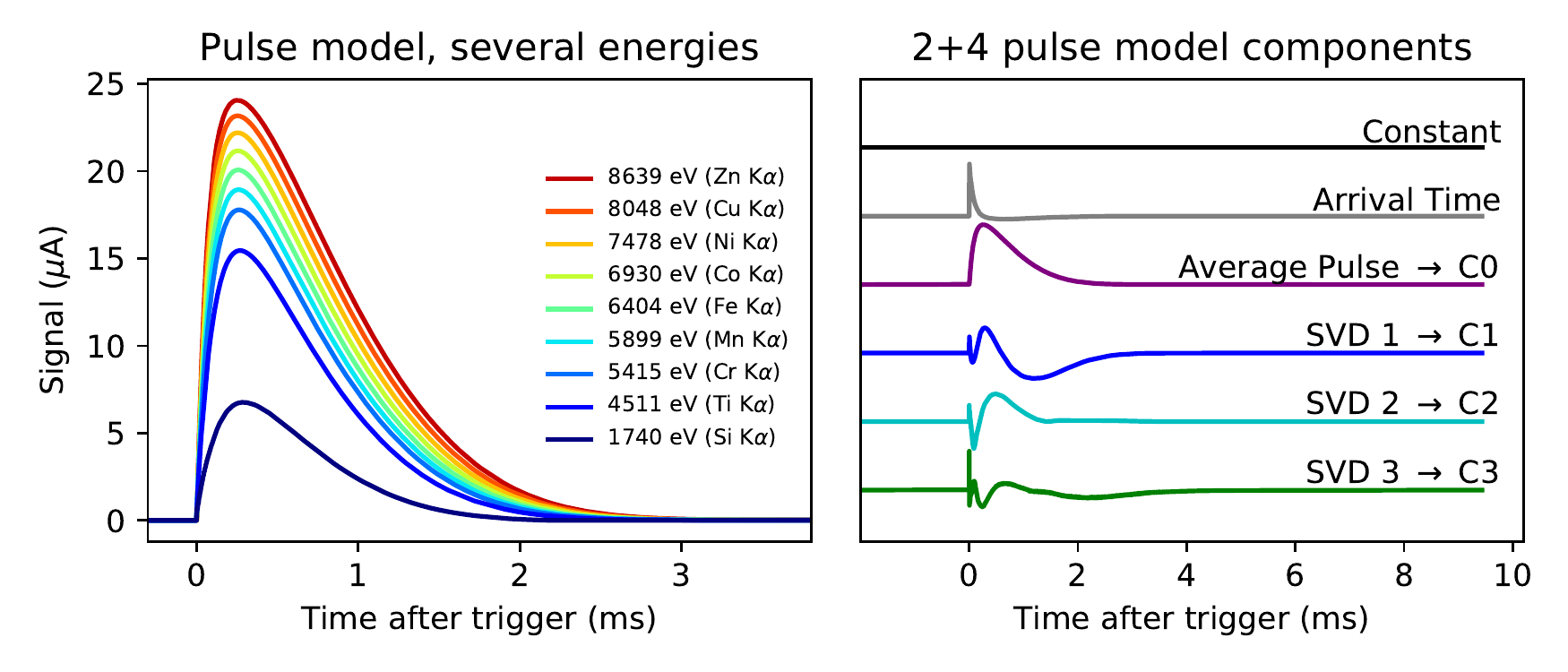}
\caption{\label{fig:svd}
\emph{Left:} The pulse model at several K$\alpha$ emission energies. \emph{Right:} Components that make up the linear subspace into which pulses are projected: three from the usual optimal filter analysis and three more from the SVD of residuals after fitting for the first three. Three singular vectors were chosen, because three singular values exceed the arbitrary cutoff of $10^{-3}$ times the largest singular value.
(Color figure online.)}
\end{center}
\end{figure}

In the usual optimal-filtering approach,\cite{fowler_ppp} we model pulse records as the sum of three components: a constant baseline level; a term that represents the linear-order correction to pulses for their exact arrival time; and an ``average pulse.'' The best-fit contribution of the last term to a record is taken to be the \emph{optimal filtered pulse height}. To find a subspace that accounts more fully for how pulse shapes vary with energy, we fit all training pulses as a sum of these three standard components, then apply singular-value decomposition (SVD) to the residuals. Fig.~\ref{fig:svd} shows the standard components and the residuals' leading singular vectors. These constitute the 6-dimensional linear subspace into which all pulses will be projected. The projection should minimize Mahalanobis (signal-to-noise) instead of Euclidean distance, in order to preserve the highest possible energy resolution.
The SVD of residuals is only one of many possible ways to choose a useful subspace of low dimension, and it is a subject of continuing research to find a procedure that is generally applicable and minimally susceptible to noise in the training data.

\begin{figure}[htbp]
\begin{center}
\includegraphics[width=\linewidth, keepaspectratio]{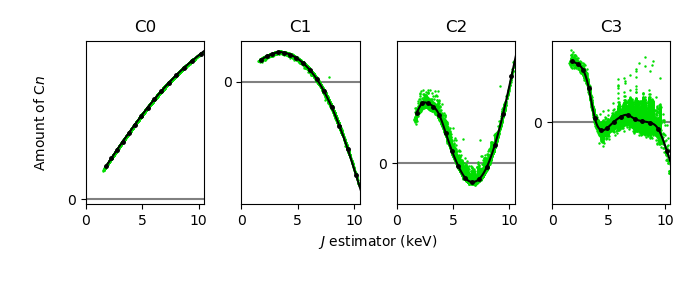}
\caption{\label{fig:nonlin_model}
The best-fit contribution of each component C0--C3 (see Fig.~\ref{fig:svd}) as a function of the pulse records' initial estimate of $J$. Small dots indicate individual training pulses; the curves connecting the larger dots indicate the spline model and its knots.
(Color figure online.)}
\end{center}
\end{figure}

In the 6-dimensional subspace, the first two dimensions are ``nuisance'' components (a constant offset and arrival-time correction); they tell us nothing about the pulse energy. The relevant information is the contribution of the other four components to each pulse record. To find the curve in this 4-dimensional space defined by the training data, 
we plot each of the four versus the (noisy) $J$ integral for each pulse, as shown in Fig.~\ref{fig:nonlin_model}. We produce a cubic-spline model for each component as a function of $J$  (Equation~\ref{eq:J}). The four splines together define a 1-dimensional curve in the 4-dimensional linear subspace; ideally all pulse records would lie on this curve, were it not for noise.

\begin{figure}[htbp]
\begin{center}
\includegraphics[width=\linewidth, keepaspectratio]{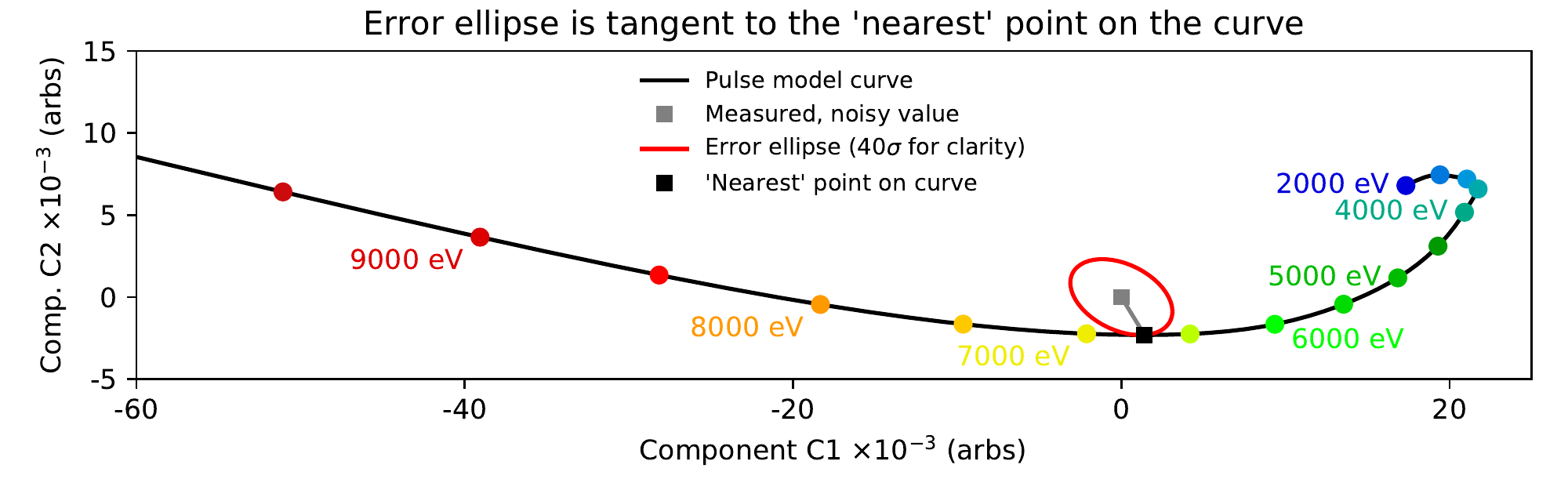}
\caption{\label{fig:nearest_point}
Graphically, the minimization of Mahalanobis distance is equivalent to finding the smallest error ellipse centered on the measurement (\emph{gray square}) that is tangent to the curve of ideal noise-free pulses. The energy is then the value of $J$ for the point (\emph{dark square}) where the ellipse is tangent to the curve. (Color figure online.)}
\end{center}
\end{figure}

The steps described so far make use of training data, which might be all available data or only a subset (say, the first $10^3$--$10^4$ clean pulse records). They define the $J$-parameterized curve on which we expect all hypothetical, noise-free pulses to lie. Because of noise, an actual measurement will lie near, but not on, the curve. To estimate $J$ for any measured pulse, we must find the value of $J$ for the nearest point on the curve. As with the linear step (the projection into a 6-dimensional subspace), we must again define ``nearest'' to minimize Mahalanobis distance (see Fig.~\ref{fig:nearest_point}). This is most readily accomplished by a noise-whitening transformation of the signal subspace,\cite{fowler_mpf} followed by a 1-dimensional minimization of the Euclidean distance between the whitened data record and the ideal-pulse curve.

\section{An Application to TES Measurements of K$\alpha$ Lines}

\begin{figure}[tbp]
\begin{center}
\includegraphics[width=.94\linewidth, keepaspectratio]{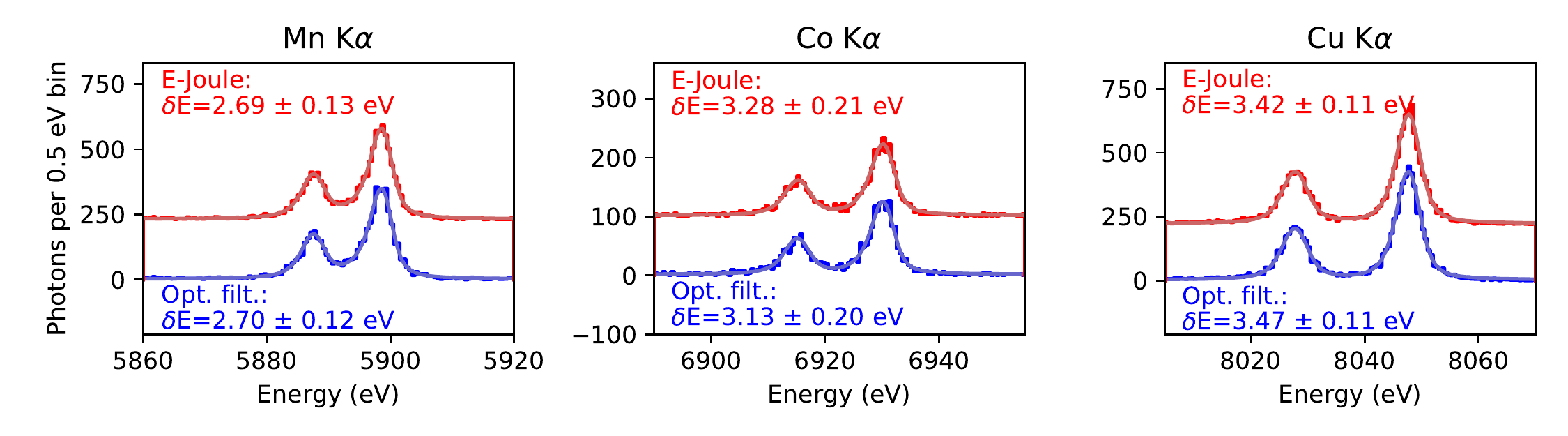}
\hspace*{.02\linewidth}\includegraphics[width=\linewidth, keepaspectratio]{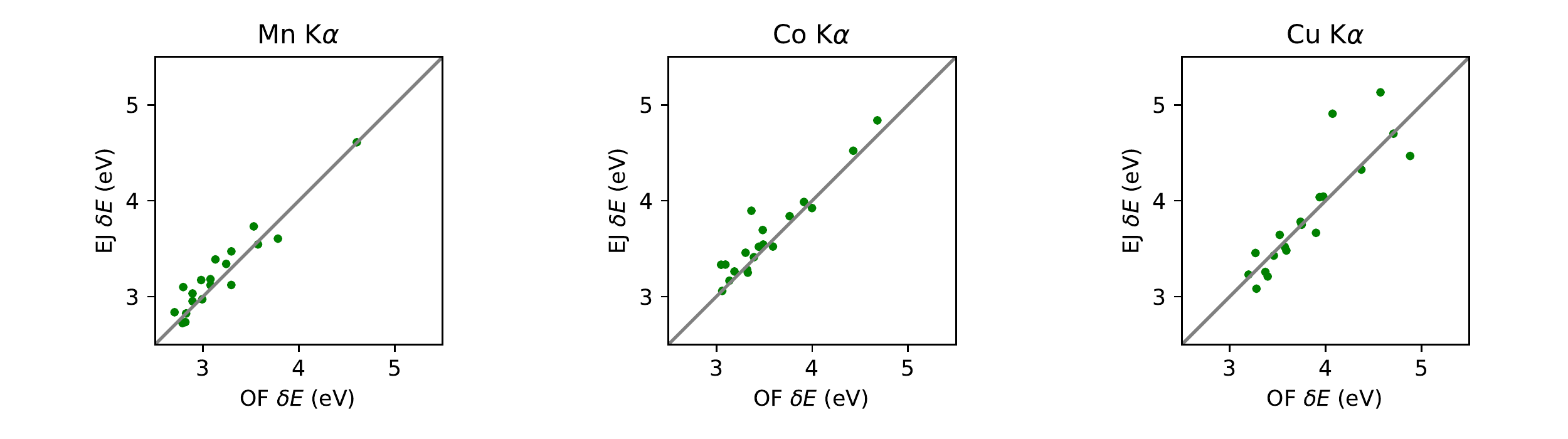}
\caption{\label{fig:resolutions}
\emph{Top:} The energy resolution for the $E_\mathrm{Joule}$ estimator $J$ and for the  optimal filter for one TES at the Mn, Co, and Cu K$\alpha$ lines. \emph{Bottom:} The resolution for the two estimators compared at the same K$\alpha$ lines, with each point representing one TES. Four TESs operated with minimal inductance have large arrival-time systematic errors and are not included.
(Color figure online.)}
\end{center}
\end{figure}

\begin{figure}[tbp]
\begin{center}
\includegraphics[width=.95\linewidth, keepaspectratio]{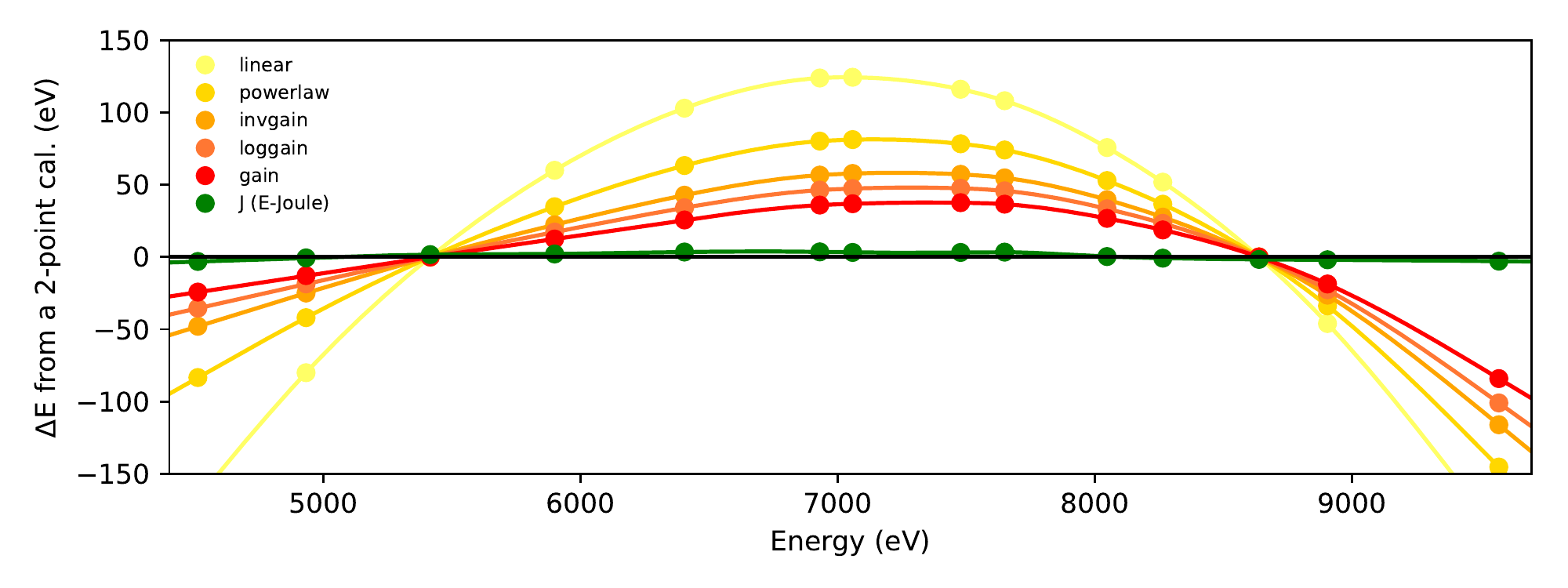}
\caption{\label{fig:linearity}
The difference between energy estimated by various methods that use precisely two free parameters and fully calibrated energy. From top to bottom at 7\,keV, the first five curves use optimal filters and assume that energy is linear or a power-law in pulse height, or that $1/G$, $\log G$, or $G$ is linear in pulse height. These curves all meet two constraints: they agree with the data exactly at the Cr and Zn K$\alpha$ energies (5415\,eV, 8639\,eV). Points indicate the K$\alpha$ and K$\beta$ emission lines; curves are splines between these points.The last curve shows that the $J$ estimator---which also has two free parameters, $\lambda$ and $\sigma$---is far more linear.
 (Color figure online.)}
\vspace{-2mm}
\end{center}
\end{figure}

We have applied this method to a data set in which a detector test array of 23 active TESs was illuminated with an x-ray source consisting of several 3d transition metals (Ti, Mn, Fe, Co, Ni, Cu, and Zn), made to fluoresce by a commercial x-ray tube source. The two-peaked K$\alpha$ complexes are useful for the assessment of the energy resolution in the 4.5\,keV to 8\,keV range. Fig.~\ref{fig:resolutions} compares the resolutions between the results of standard optimal filtering and $J$ as estimated by the nearest-point-on-the-curve procedure. The energy resolutions are similar, typically 2.7\,eV to 4\,eV FWHM at 6\,keV, while the nonlinearity of the $J$ estimator (Fig.~\ref{fig:linearity}) is an order of magnitude smaller.

\section{Future Prospects and Conclusion}

In the present data set, the Joule-energy estimator exhibits an energy resolution consistent with that provided by optimal filtering. Even a simpler, noisier $J$ estimator than this one could serve as a useful complement to linear optimal filtering, offering an energy pre-calibration. An additional correction for residual nonlinearity would be necessary, but it would be more constrained than the conversion  from optimally filtered pulse heights to energy. We have previously found uncertainty in the nonlinear response of a TES microcalorimeter to be the leading source of systematic error on the absolute energy scale.\cite{fowler17} 
The nonlinear estimator $J$ was studied to determine whether it can reduce these systematic errors, but the present data set is unfortunately too small for a full validation of the calibration. 
Of course, a pre-calibration requiring only two free parameters and accurate to one part in 10$^3$ would be very convenient, separate from any reduction in the final systematic errors. The integral-based approach is also flexible: Eq.~\ref{eq:J} could be extended with integrals of higher powers of $s(t)$ if they proved useful.

The low-noise estimation of $J$ also serves as an example of how one might perform any nonlinear---but statistically optimal---analysis of microcalorimeter pulses. In our approach to this problem, the computational burden of the nonlinear estimator is only a constant factor larger than that required by optimal filtering, regardless of the pulse-record length. After the analysis of training data, the dominant computational step performed on each pulse record is its projection into a low-dimensional subspace. This is a linear operation that requires only inner products of the record with one constant weighting vector for each dimension in the linear subspace. This step drastically reduces the size of the data vector (here, from $N$ time samples to 6 coordinates in the subspace). All further processing, even if it requires iterative steps (e.g., to find the nearest point on the good-pulse curve), is performed on data of the reduced size and is unlikely to dominate the computation. Therefore, nonlinear optimal analysis of this type will be possible even a highly resource-constrained environment such as a spacecraft.

\begin{acknowledgements}
This work was supported by NIST's Innovations in Measurement Science program and by NASA SAT NNG16PT18I, ``Enabling \& enhancing technologies for a demonstration model of the Athena X-IFU.'' C.G.P. is supported by a National Research Council Post-Doctoral Fellowship. The contribution of NIST is not subject to copyright.
\end{acknowledgements}


\end{document}